\documentclass[showpacs,onecolumn,pra]{revtex4}
\usepackage{amsmath}
\usepackage{amssymb}
\usepackage{graphicx}
\usepackage{subfigure}

\begin{document}

\title{Spontaneous symmetry breaking in a nonlinear double-well structure}
\author{Thawatchai Mayteevarunyoo$^{1,2}$, Boris A. Malomed$^{2}$, and
Guangjiong Dong$^{3}$}
\affiliation{$^{1}$Department of
Telecommunication Engineering, Mahanakorn University of
Technology, Bangkok 10530, Thailand\\
$^{2}$Department of Physical Electronics, School of Electrical
Engineering,
Faculty of Engineering, Tel Aviv University, Tel Aviv 69978, Israel\\
$^{3}$State Key Laboratory of Precision Spectroscopy and Physics
Department, East China Normal University, Shanghai, China}

\begin{abstract}
We propose a model of a nonlinear double-well potential (NDWP), alias a
double-well pseudopotential, with the objective to study an alternative
implementation of the spontaneous symmetry breaking (SSB) in Bose-Einstein
condensates (BECs) and optical media, under the action of a potential with
two symmetric minima. In the limit case when the NDWP structure is induced
by the local nonlinearity coefficient represented by a set of two
delta-functions, a fully analytical solution is obtained for symmetric,
antisymmetric and asymmetric states. In this solvable model, the SSB
bifurcation has a fully subcritical character. Numerical analysis, based on
both direct simulations and computation of stability eigenvalues,
demonstrates that, while the symmetric states are stable up to the SSB
bifurcation point, both symmetric and emerging asymmetric states, as well as
all antisymmetric ones, are unstable in the model with the delta-functions.
In the general model with a finite width of the nonlinear-potential wells,
the asymmetric states quickly become stable, simultaneously with the switch
of the SSB bifurcation from the subcritical to supercritical type.
Antisymmetric solutions may also get stabilized in the NDWP structure of the
general type, which gives rise to a bistability between them and asymmetric
states. The symmetric states require a finite norm for their existence, an
explanation to which is given. A full diagram for the existence and
stability of the trapped states in the model is produced. Experimental
observation of the predicted effects should be possible in BEC formed by
several hundred atoms.
\end{abstract}

\maketitle

\section{Introduction}

The one-dimensional (1D) Schr\"{o}dinger equation including a symmetric
potential structure produces single-particle wave functions of a definite
parity, even or odd, with the ground state always corresponding to an even
function without zeros. However, a spatially symmetric Hamiltonian of an
interacting many-particle system can give rise to asymmetric states, which
may be considered as a spontaneous-symmetry-breaking (SSB) effect. At the
classical level, the SSB occurs in optics, as a result of the interplay
between the nonlinearity and waveguiding structures, when the strong
nonlinearity partly suppresses the linear coupling between parallel guiding
cores. In particular, it was shown that a stable trapped mode may be
asymmetric in a channel waveguide embedded in the self-focusing Kerr medium
\cite{chan}. The onset of a sharp symmetry-breaking instability in a
double-hump two-component spatial optical soliton was demonstrated
experimentally in a planar nonlinear waveguide \cite{ob}.

A natural setting in which SSB phenomena may arise in the context of
nonlinear optics and Bose-Einstein condensation (BEC) is provided by
double-well potentials (DWPs). In the experiment, an effective optical DWP
was created by a specially designed illumination pattern applied, in the
ordinary polarization, to a photorefractive crystal (the SSB was observed in
a beam with extraordinary polarization, shone through this structure) \cite%
{Zhigang}. It was also proposed to realize similar effective potentials in
coupled nonlinear microcavities \cite{Maes}, and in a structured core of an
optical fiber \cite{Longhi}. A specific variety of the optical SSB was
studied in a model of two parallel-coupled antiwaveguides with the
self-focusing nonlinearity, which corresponds to an effective double-barrier%
\textit{\ }potential, rather than DWP \cite{Kaplan}.

Well-known dual-core optical fibers \cite{dual-core}, which may serve as a
basis for the power-controlled all-optical switching, if the Kerr
nonlinearity is taken into regard \cite{switch}, may also be considered as
DWP structures, with the difference that the tunneling between two potential
wells is replaced by the linear coupling between the cores. In addition to
the SSB of continuous-wave states \cite{Canberra}, the formation of
asymmetric solitons in dual-core fibers was studied in detail theoretically
\cite{dual-core-fiber}. Similar analysis of the SSB for soliton modes was
performed in models of dual-core fiber Bragg gratings with the Kerr
nonlinearity \cite{dual-core-FBG}, and coupled waveguides with the quadratic
\cite{dual-core-quadratic} and cubic-quintic \cite{dual-core-CQ} nonlinear
terms, including a system of linearly coupled complex Ginzburg-Landau
equations of the cubic-quintic type \cite{Sigler}. In Refs. \cite{Buryak},
\cite{Arik} and \cite{Skryabin},\ the analysis of the SSB was extended to
three-core linearly coupled triangular configurations -- for optical fibers,
Bragg gratings, and complex Ginzburg-Landau equations, respectively.

The concept of SSB also plays an important role in understanding
experimental phenomena in BEC, because, if interactions between atoms are
strong enough, the ground state of the condensate may not follow the
symmetry of the trapping potential \cite{esr}. In particular, manifestations
of SSB were observed in a quenched ferromagnetic state of a spinor
(three-component) condensate \cite{sad}. In the single-component BEC, a
natural setting for the realization of SSB may again be provided by DWP
configurations. An effectively one-dimensional DWP structure was realized
experimentally in Ref. \cite{markus1}. Loading a condensate of $^{87}$Rb
atoms with the repulsive interaction between them into this structure made
it possible to observe Josephson oscillations for a small number of atoms,
and the macroscopic quantum self-trapping featuring an imbalance between
populations of the two wells, for a larger number. Parallel to the
experimental work, numerous theoretical studies of matter-wave DWP\ settings
have been performed, for the cases of both repulsion and attraction between
atoms. These studies addressed problems such as finite-mode reductions \cite%
{finite-mode} (including two-component mixtures \cite{Amherst}), obtaining
analytical results for specific shapes of the potential \cite{exact},
quantum effects \cite{quantum}, and some others. Recently investigated
tunneling between vortex and antivortex states in BEC trapped in a 2D
anisotropic potential \cite{Pethick} belongs to this category too.

Theoretical analysis was also performed for 2D and 3D extensions of the DWP
settings in BEC, which add one or two extra dimensions to the model, either
without an additional potential, or with a periodic optical-lattice (OL)
potential acting in these directions. These settings may be approximated,
similar to the above-mentioned standard model of dual-core optical fibers,
by a system of linearly coupled 1D \cite{2DArik} or 2D \cite{3D} equations.
In a more accurate form, nearly-1D solitons can be found as solutions to the
full 2D equation that includes the DWP (the potential depends on the
transverse coordinate, $x$, allowing solitons to self-trap in the free
longitudinal direction, $y$) \cite{2DMichal}. The latter model is relevant
to the case of the self-attractive nonlinearity. In the case of
self-repulsion, dual-core gap solitons have been be predicted in the setting
with the OL potential applied along direction $y$ \cite{MarekMarkus}. Note
that, in any setting, gap solitons cannot realize the ground state of the
respective system, but, nevertheless, they represent stable configurations,
that have been created in the experiment using the condensate of $^{87}$Rb
atoms with the repulsion between them \cite{Heidelberg}.

A general principle, upheld by the analysis in various settings (in
nonlinear optics and BEC alike), is that the SSB occurs through bifurcations
of symmetric or antisymmetric states, in the models with the self-attraction
and self-repulsion, respectively. As mentioned above, models of the
DWP/double-core type, combining cubic attractive and quintic repulsive
nonlinearities, were studied too \cite{dual-core-CQ,Sigler,Radik,Zeev}. In
the latter case, the competition between the self-focusing and
self-defocusing against the backdrop of the DWP structure gives rise to
specific SSB bifurcation diagrams, in the form of non-convex closed loops
\cite{dual-core-CQ,Zeev}, as well as to specific dynamical switching regimes
\cite{Radik}. Also predicted were manifestations of the SSB in a
two-component BEC\ mixture trapped in the DWP structure \cite{Amherst}, for
both cases of the self-attraction and self-repulsion.

All the extensive work on the SSB outlined above was performed in settings
based on usual linear potentials of the double-well type. The objective of
the present work is to propose another physical framework, in which the SSB
can be predicted in an effective \emph{nonlinear} double-well potential
(NDWP), induced through a spatial modulation of the local nonlinearity
coefficient. Following the terminology commonly adopted in the solid-state
theory \cite{Harrison}, this nonlinear ingredient of the physical model may
also be a called a \textit{pseudopotential}.

In BEC settings, a pseudopotential structure may be readily induced through
spatial modulation of the local value of the $s$-wave scattering length, $%
a_{s}(x)$, which determines the effective BEC\ nonlinearity. The modulation
can be implemented, through the Feshbach resonance, by means of a spatially
inhomogeneous dc magnetic field \cite{FR-magnetic}, or by a resonant optical
field, as predicted in Ref. \cite{FR-optical} and demonstrated
experimentally in Ref. \cite{FR-optical-exp}. It was also proposed to
control the Feshbach resonance by dint of dc electric field \cite%
{FR-electric}, which can be easily made inhomogeneous too. The attractive
and repulsive interactions between atoms correspond to $a_{s}<0$ and $%
a_{s}>0 $, respectively; both signs, as well as sign-changing patterns, can
be used to engineer effective nonlinear potentials.

So designed pseudopotential lattices have attracted much interest in studies
of BEC. In the 1D geometry, solitons, extended wave patterns, and various
dynamical states supported by such structures were studied theoretically
\cite{nonlin-periodic,Panos} (a random nonlinear lattice \cite{random} and
pseudopotentials generated by a spatially monotonous ramp of the local
scattering length \cite{ramp} were explored too). Recently, similar states
were also considered in nonlinear optics, assuming a periodic modulation of
the local Kerr coefficient \cite{nonlin-periodic-optics}. Some (but much
fewer) results were obtained too for 2D settings \cite{2Dnonlin}.

However, to the best of our knowledge, SSB phenomena in nonlinear
pseudopotentials have not been studied yet. In this work, we focus on such
effects in NDWP settings, which can be engineered by means of techniques
mentioned above, using attractive interactions between atoms in BEC, or the
self-focusing nonlinearity in optics, as briefly described below. In Section
II, we formulate the model and give estimates of characteristic values of
related physical parameters. Section III reports full analytical solutions
corresponding to symmetric, antisymmetric, and \emph{asymmetric} states
trapped by the NDWP, in the limit case with the modulation of the local
nonlinearity coefficient is represented by a set of two Dirac's
delta-functions. The relevance of the latter model is stressed, in
particular, by the recently introduced \cite{Panos} BEC model with a
periodic nonlinear potential of the Kronig-Penney type, whose simplest
version reduces to a periodic array of delta-functions (SSB effects were not
studied in Ref. \cite{Panos}).

In section IV, we present numerical results for the general model, in which
the delta-functions are replaced by a pair of Gaussians of a finite width.
In that case, the trapped states are found in a numerical form, and their
stability is studied by means of direct simulations of slightly perturbed
stationary states and also, independently, through computation of respective
stability eigenvalues for small perturbations. The result is that asymmetric
states, which are unstable in the delta-function limit, can be readily
stabilized in the general model. In addition, antisymmetric states may be
stabilized too, in two disjoint regions of the parameter state, giving rise
to a \textit{bistability} involving antisymmetric and asymmetric states. A
condition for the existence of symmetric states is that their norm must
exceed a certain threshold value, in terms of their norm, an explanation to
which is given. The existence and stability of all states, including a line
of the SSB bifurcation, are summarized in a single diagram, which is
presented in Section IV too. Results reported in this paper and perspectives
for further work are summarized in Section V.

\section{The model}

The underlying 3D Gross-Pitaevskii equation for the mean-field wave
function, $\Psi (X,Y,Z,T)$, is taken in the ordinary form corresponding to
the nearly-1D trap \cite{pit}:%
\begin{equation}
i\hbar \Psi _{T}=-\frac{\hbar ^{2}}{2m}\nabla ^{2}\Psi +\frac{m\omega
_{\perp }^{2}}{2}R^{2}\Psi +\frac{4\pi \hbar ^{2}a_{s}(X)}{m}\left\vert \Psi
\right\vert ^{2}\Psi ,  \label{GPE}
\end{equation}%
with $m$ the atomic mass and $\omega _{\perp }$ the frequency providing for
the tight confinement of the condensate in the direction of $R\equiv \sqrt{%
Y^{2}+Z^{2}}$. As said above, we assume an axial modulation (and negative
sign) of the scattering length, in the form of a pair of Gaussians, each of
width $l$ and amplitude $A_{0}$, with centers set at points $X=\pm \Lambda $%
:
\begin{equation}
a_{s}(X)=-A_{0}\left[ \exp \left( -\frac{\left( X+\Lambda \right) ^{2}}{l^{2}%
}\right) +\exp \left( -\frac{\left( X-\Lambda \right) ^{2}}{l^{2}}\right) %
\right] .  \label{Gauss}
\end{equation}%
In what follows below, we measure the axial coordinate in units of $\Lambda $%
, and, accordingly, time in units of $m\Lambda ^{2}/\hbar $, i.e., we define
\begin{equation}
x\equiv X/l,a\equiv l/\Lambda ,t\equiv \hbar T/\left( m\Lambda ^{2}\right) .
\label{a}
\end{equation}%
Further, following the usual approach to the derivation of the effective 1D
equation, we take 3D field as a product of a slowly varying axial function, $%
\psi (X,T)$, and the ground-state wave function in the transverse plane,
\begin{equation}
\Psi =\frac{1}{\sqrt{2\pi ^{5/2}aA_{0}\Lambda ^{2}}}\exp \left( i\omega
_{\perp }T-\frac{R^{2}}{2a_{\perp }^{2}}\right) \psi (x,t),  \label{Psi}
\end{equation}%
with $a_{\perp }^{2}\equiv \hbar /\left( m\omega _{\perp }\right) $ [the
scaling factor here is chosen so as to maintain normalization condition (\ref%
{scaling}), see below].

\begin{figure}[h]
\centering\includegraphics[width=3.5in]{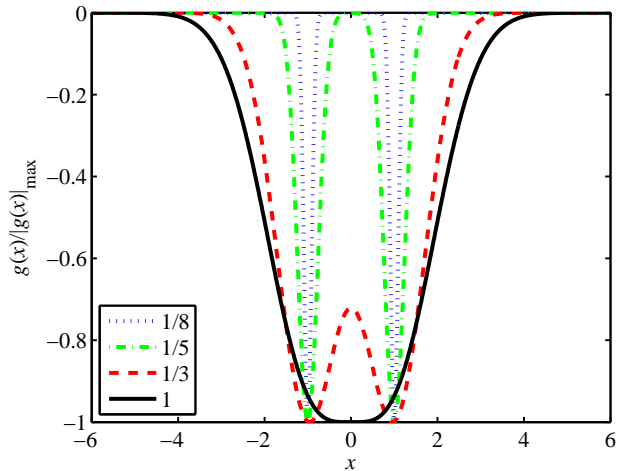}
\caption{(Color online) Shapes of double-well function (\protect\ref{g}),
normalized to its maximum value, are shown for different values of scaled
width $a$ of the individual well.}
\label{Fig1}
\end{figure}

The substitution of expression (\ref{Psi}) in Eq. (\ref{GPE}) and averaging
in the transverse plane lead to the following 1D equation:%
\begin{equation}
i\psi _{t}=-\frac{1}{2}\psi _{xx}+g(x)|\psi |^{2}\psi ,  \label{psi2}
\end{equation}%
\begin{equation}
g(x)=-\frac{1}{a\sqrt{\pi }}\left[ \exp \left( -\frac{\left( x+1\right) ^{2}%
}{a^{2}}\right) +\exp \left( -\frac{\left( x-1\right) ^{2}}{a^{2}}\right) %
\right] ,  \label{g}
\end{equation}%
where the modulated nonlinearity coefficient is subject to the following
normalization condition:
\begin{equation}
\int_{-\infty }^{+\infty }g(x)dx~\equiv 2.  \label{scaling}
\end{equation}%
Profiles of modulation function (\ref{g}), which keeps the double-well
structure for $a<\sqrt{2}$, are shown, for different values of $a$, in Fig. %
\ref{Fig1}.

Equation (\ref{psi2}) conserves two dynamical invariants, \textit{viz}., the
norm and energy (Hamiltonian),
\begin{equation}
N\equiv \int_{-\infty }^{+\infty }|\psi (x)|^{2}dx,~H=\frac{1}{2}%
\int_{-\infty }^{\infty }\left( \left\vert \psi _{x}\right\vert
^{2}+g(x)\left\vert \psi \right\vert ^{4}\right) dx.  \label{N}
\end{equation}%
As follows from Eqs. (\ref{Psi}) and (\ref{a}), the total number of atoms in
the condensate, $\mathcal{N}$, is related to 1D norm $N$ as follows:%
\begin{equation}
\mathcal{N}\equiv \int \int \int \left\vert \Psi \left( X,Y,Z\right)
\right\vert ^{2}dXdYdZ=\frac{a_{\perp }^{2}}{2\pi ^{3/2}aA_{0}\Lambda }N%
\text{.}  \label{number}
\end{equation}%
For physical parameters relevant to experiments with the condensate of $^{7}$%
Li atoms \cite{RH}, i.e., $a_{\perp }\sim 2~\mu $m, $A_{0}\sim 0.5$ nm, and $%
\Lambda \sim 20$ $\mu $m, characteristic values of the number of atoms in
various patterns reported below [see, in particular, Figs. \ref{Fig4}(d) and %
\ref{Fig8}] fall into the range of $\mathcal{N}$ between $\sim 200$ and $%
1000 $, which is quite sufficient for experimental manipulations and
observation of the patterns. In the same range of physical parameters, $t=1$
in Eq. (\ref{psi2}), is estimated as being tantamount to $\sim 10$ ms, hence
typical time scales for the instability development or intrinsic
oscillations of breathers induced by the instabilities, that are reported
below, are expected to be in the range of $0.1$ to $1$ s, which is realistic
to the currently available experimental techniques \cite{RH}.

In terms of optical settings, a set of two narrow (of width $l\sim 1$ $\mu $%
m) parallel stripes with strong local nonlinearity can be built, in a planar
waveguide, by means of available nanotechnological methods. In that case,
the power of the laser beam necessary for the self-trapping of transverse
nonlinear patterns in the waveguide made of silica may be $\sim 500$ kW,
\cite{Silberberg} while, using AlGaAs, one may reduce the necessary power to
the level of $1$ kW \cite{PhysRep}. In these settings, the characteristic
evolution length of the spatial beam can be made shorter than $1$ mm.
Obviously, transitions between states of different types reported in this
paper may be relevant to the design of power-controlled optical-switching
schemes. On the other hand, the description of the planar waveguide with the
pair of embedded stripes may require a model more general than the one
studied here, as it will plausibly combine the transverse modulation of the
local nonlinearity with a similar linear potential (which is briefly
described at the end of the next section), as the material difference
between the stripes and host medium ought to affect the linear index of
refraction too.

Stationary localized solutions to Eq. (\ref{psi2}) are sought for as $\psi
=e^{-i\mu t}\phi (x)$, where the chemical potential is negative, $\mu <0$,
and real function $\phi (x)$ satisfies equation%
\begin{equation}
\mu \phi +\frac{1}{2}\phi ^{\prime \prime }+\frac{1}{a\sqrt{\pi }}\left[
\exp \left( -\frac{\left( x+1\right) ^{2}}{a^{2}}\right) +\exp \left( -\frac{%
\left( x-1\right) ^{2}}{a^{2}}\right) \right] \phi ^{3}=0.  \label{phi2}
\end{equation}

An analytically solvable version of the model corresponds to the limit of $%
a\rightarrow 0$, with Eq. (\ref{psi2}) going over into%
\begin{equation}
i\psi _{t}=-(1/2)\psi _{xx}-\left[ \delta (x+1)+\delta (x-1)\right] |\psi
|^{2}\psi ,  \label{psi}
\end{equation}%
where $\delta (x)$ is the Dirac's delta-function. Note that rescaled
equation (\ref{psi}) contains no free parameters; however, the norm of the
solution will play the role of an intrinsic parameter, see below. In the
same limit, stationary equation (\ref{phi2}) takes the form of%
\begin{equation}
\mu \phi +(1/2)\phi ^{\prime \prime }+\left[ \delta (x+1)+\delta (x-1)\right]
\phi ^{3}=0.  \label{phi}
\end{equation}

\section{The model with the delta-functions: analytical solutions}

\subsection{Symmetric, antisymmetric, and asymmetric states}

Off points $x=\pm 1$, Eq. (\ref{phi}) is linear. A general solution to this
equation, decaying at $|x|\rightarrow \infty $, can be written as%
\begin{equation}
\phi (x)=\left\{
\begin{array}{c}
B_{1}e^{\sqrt{2|\mu |}\left( x+1\right) },~\mathrm{at}~x<-1, \\
A_{0}e^{-\sqrt{2|\mu |}\left( x-1\right) }+B_{0}e^{\sqrt{2|\mu |}\left(
x+1\right) },~\mathrm{at}~-1<x<+1, \\
A_{1}e^{-\sqrt{2|\mu |}\left( x-1\right) },~\mathrm{at}~x>+1,%
\end{array}%
\right.  \label{linear}
\end{equation}%
with constant amplitudes $A_{0},A_{1}$ and $B_{0},B_{1}$. The continuity of
the wave function at $x=\pm 1$ imposes two relations on them, $%
B_{1}=B_{0}+A_{0}e^{2\sqrt{2|\mu |}},~A_{1}=A_{0}+B_{0}e^{2\sqrt{2|\mu |}},$
which allows one to eliminate $A_{0}$ and $B_{0}$ in favor of $A_{1}$ and $%
B_{1}$,
\begin{equation}
A_{0}=\frac{e^{2\sqrt{2|\mu |}}B_{1}-A_{1}}{e^{4\sqrt{2|\mu |}}-1},~B_{0}=%
\frac{e^{2\sqrt{2|\mu |}}A_{1}-B_{1}}{e^{4\sqrt{2|\mu |}}-1}.  \label{A0B0}
\end{equation}%
Further, the integration of Eq. (\ref{phi}) in infinitesimal vicinities of
points $x=\pm 1$ yields expressions for jumps ($\Delta $) of the first
derivative at these points, $\Delta \left( \phi ^{\prime }\right) |_{x=\pm
1}=-2\left( \phi |_{x=\pm 1}\right) ^{3}$. The substitution of solution (\ref%
{linear}) in these relations leads to a system of cubic equations for the
amplitudes,%
\begin{eqnarray}
\sqrt{|\mu |/2}\left( B_{1}-B_{0}+A_{0}e^{2\sqrt{2|\mu |}}\right)
&=&B_{1}^{3},  \label{cubicA} \\
\sqrt{|\mu |/2}\left( A_{1}-A_{0}+B_{0}e^{2\sqrt{2|\mu |}}\right)
&=&A_{1}^{3}.  \label{cubicB}
\end{eqnarray}%
After the substitution of expressions (\ref{A0B0}) into Eqs. (\ref{cubicB})
and (\ref{cubicA}), we end up with two coupled cubic equations for $A_{1}$
and $B_{1}$:%
\begin{equation}
\sqrt{2|\mu |}e^{2\sqrt{2|\mu |}}\left( e^{2\sqrt{2|\mu |}%
}B_{1}-A_{1}\right) =\left( e^{4\sqrt{2|\mu |}}-1\right) B_{1}^{3},
\label{A}
\end{equation}%
\begin{equation}
\sqrt{2|\mu |}e^{2\sqrt{2|\mu |}}\left( e^{2\sqrt{2|\mu |}%
}A_{1}-B_{1}\right) =\left( e^{4\sqrt{2|\mu |}}-1\right) A_{1}^{3}.
\label{B}
\end{equation}

Solving Eqs. (\ref{A}) and (\ref{B}), we first find symmetric and
antisymmetric solutions,
\begin{equation}
A_{1}=B_{1}\equiv A_{\mathrm{sym}}=\pm \sqrt{\frac{\sqrt{2|\mu |}}{1+e^{-2%
\sqrt{2|\mu |}}}}.  \label{symm}
\end{equation}%
\begin{equation}
A_{1}=-B_{1}\equiv A_{\mathrm{antisym}}=\pm \sqrt{\frac{\sqrt{2|\mu |}}{%
1-e^{-2\sqrt{2|\mu |}}}}.  \label{antisymm}
\end{equation}%
The norm of these solutions, defined as per Eq. (\ref{N}), is

\begin{equation}
N_{\mathrm{sym,antisym}}=\frac{1}{1\pm e^{-2\sqrt{2|\mu |}}}+\frac{1-e^{-4%
\sqrt{2|\mu |}}\pm 4\sqrt{2|\mu |}e^{-2\sqrt{2|\mu |}}}{\left( 1\pm e^{-2%
\sqrt{2|\mu |}}\right) ^{3}},  \label{+-}
\end{equation}%
with $+$ and $-$ corresponding to the symmetric and antisymmetric states,
respectively. In the limit of $\mu \rightarrow -\infty ,$ both expressions (%
\ref{symm}) and (\ref{antisymm}) yield $A_{\mathrm{sym}}^{2}\left( \mu
=-\infty \right) =A_{\mathrm{antisym}}^{2}\left( \mu =-\infty \right) =\sqrt{%
2|\mu |}$. In this limit, norm (\ref{+-}) of both solutions takes a common
value, $N_{\mathrm{sym,antisym}}\left( \mu =-\infty \right) =2.$On the other
hand, in the limit of $\mu \rightarrow -0$, the amplitude of the symmetric
state vanishes, $A_{\mathrm{sym}}(\mu \rightarrow -0)\approx \left( |\mu
|/2\right) ^{1/4}$, and its norm takes a finite limit value, $N_{\mathrm{sym}%
}(\mu =0)=1/2$, while the amplitude of the antisymmetric state remains
finite, $A_{\mathrm{antisym}}(\mu \rightarrow -0)=1/\sqrt{2}$, and its norm
diverges, $N_{\mathrm{antisym}}(\mu \rightarrow -0)\approx 1/\left( 2\sqrt{%
2|\mu |}\right) $. It can be easily checked that the decrease of $N_{\mathrm{%
sym}}$ and increase of $N_{\mathrm{antisym}}$ with the variation of $\mu $
from $-\infty $ to $0$ are not monotonous: $N_{\mathrm{sym}}$ attains a
maximum, $\left( N_{\mathrm{sym}}\right) _{\max }\approx 2.08$, at $\mu
\approx -1.40$, and $N_{\mathrm{antisym}}$ has a minimum, $\left( N_{\mathrm{%
antisym}}\right) _{\min }\approx 1.84$, at $\mu \approx -0.58$.

Actually, the finite minimum norm (\textit{threshold}), necessary for the
existence of the symmetric states, which is $N_{\mathrm{sym}}(\mu =0)\equiv
\left( N_{\mathrm{sym}}\right) _{\min }=1/2$ in the present case, is a
generic property, shared by the model with finite width $a$ of the
nonlinear-potential wells, as shown in detail below in Fig. \ref{Fig8}. On
the other hand, the existence of the above-mentioned maximum value of the
norm for the symmetric states, $\left( N_{\mathrm{sym}}\right) _{\max
}\approx 2.08$, is a specific feature of the model with $a=0$ (the one based
on the delta-functions): it is easy to see that, in the model with finite $a$%
, the norm of any state grows $\sim a\sqrt{|\mu |}$ at $\mu \rightarrow
-\infty $. However, this specific feature, which, as a matter of fact,
indicates degeneracy of the model with $a=0$, is less significant for the
physical applications, as all symmetric states, for $a=0$ and finite $a$
alike, are unstable when the norm exceeds its value at the SSB bifurcation
point [further details can be seen in Eq. (\ref{Nbif}) and Fig. \ref{Fig8}
below].

A point of the SSB bifurcation, which gives rise to a pair of asymmetric
solutions splitting off from the symmetric one, can be easily found. Indeed,
the symmetry breaking means that symmetric solution (\ref{symm}) acquires an
infinitesimal \emph{antisymmetric} addition, $\delta A_{1}=-\delta
B_{1}\equiv \delta A$. Thus, infinitely close to the bifurcation point, the
relevant solution is sought for as $A_{1}=A_{\mathrm{sym}}+\delta
A,~B_{1}=A_{\mathrm{sym}}-\delta A$. The substitution of this in Eqs. (\ref%
{A}) and (\ref{B}) and linearization in infinitesimal $\delta A$ lead to a
simple equation that predicts the value of the chemical potential at the
bifurcation point: $\exp \left( \sqrt{2\left\vert \mu _{\mathrm{bif}%
}\right\vert }\right) =\sqrt{2}$, or%
\begin{equation}
\mu _{\mathrm{bif}}=-\left( \ln 2\right) ^{2}/8\approx -0.06\,.  \label{bif}
\end{equation}%
At this point, the amplitude of symmetric solution (\ref{symm}) is $A_{%
\mathrm{bif}}=\sqrt{\left( \ln 2\right) /3}\approx 0.481$, and the value of
norm (\ref{N}), with the upper sign, is
\begin{equation}
N_{\mathrm{bif}}=2/3+(8/27)\left( 3/4+\ln 2\right) \approx 1.09.
\label{Nbif}
\end{equation}%
Actually, value (\ref{bif}) of the chemical potential at the bifurcation
point ins approximately the same in the model with finite-width
nonlinear-potential wells, up to $a\approx 1$, as seen from Fig. \ref{Fig4}%
(d) presented below.

The same analysis shows that antisymmetric solution (\ref{antisymm}) never
gives rise to an antisymmetry-breaking bifurcation. Indeed, for this
solution the antisymmetry would be broken by an infinitesimal symmetric
variation, $\delta A_{1}=\delta B_{1}\equiv \delta A$, i.e., infinitely
close to the bifurcation point, the solution would be $A_{1}=A_{\mathrm{%
antisym}}+\delta A,~B_{1}=-A_{\mathrm{antisym}}+\delta A$. Subsequent
substitution in Eqs. (\ref{A}) and (\ref{B}) and the linearization in $%
\delta A$ yield an equation for $\mu $ that has no real solutions.

\begin{figure}[h]
\centering\includegraphics[width=3.5in]{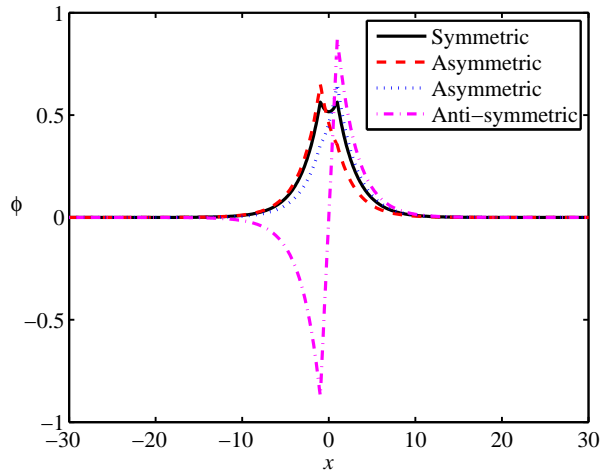}
\caption{(Color online) Profiles of symmetric, asymmetric, and antisymmetric
stationary states in the analytically solvable model with the
delta-functions, all pertaining to $\protect\mu =-0.1$. The respective
values of the norm are $N_{\mathrm{symm}}\approx 1.27$, $N_{\mathrm{asymm}%
}\approx 1.07$, and $N_{\mathrm{antisymm}}\approx 2.18$. The mutually
symmetric lines (blue and red ones, in the color version of the figure)
represent two asymmetric states that are mirror images of each other.}
\label{Fig2}
\end{figure}

Equations (\ref{A}) and (\ref{B}) can be solved analytically for asymmetric
states too:%
\begin{equation}
\left\{ A_{1},B_{1}\right\} _{\mathrm{asym}}=\frac{|\mu |^{1/4}\left( \sqrt{%
1+2e^{-2\sqrt{2|\mu |}}}\pm \sqrt{1-2e^{-2\sqrt{2|\mu |}}}\right) }{2^{3/4}%
\sqrt{1-e^{-4\sqrt{2|\mu |}}}}~.  \label{asymm}
\end{equation}%
Note that full solution (\ref{asymm}) predicts exactly the same bifurcation
point as Eq. (\ref{bif}), i.e., $\exp \left( \sqrt{2|\mu _{\mathrm{bif}}|}%
\right) =\sqrt{2}$ [at this point, the second radical in Eq. (\ref{asymm})
vanishes]. These solutions are characterized by the \textit{asymmetry ratio}%
, which is defined as
\begin{equation}
\Theta =\frac{\int_{0}^{+\infty }\phi ^{2}(x)dx-\int_{-\infty }^{0}\phi
^{2}(x)dx}{\int_{-\infty }^{+\infty }\phi ^{2}(x)dx}\equiv \frac{N_{+}-N_{-}%
}{N}.  \label{theta}
\end{equation}%
Typical examples of symmetric, asymmetric and antisymmetric states produced
by he above analytical solutions are displayed in Fig. \ref{Fig2}.

\subsection{Bifurcation diagrams and stability}

\begin{figure}[h]
\centering\subfigure[]{\includegraphics[width=3in]{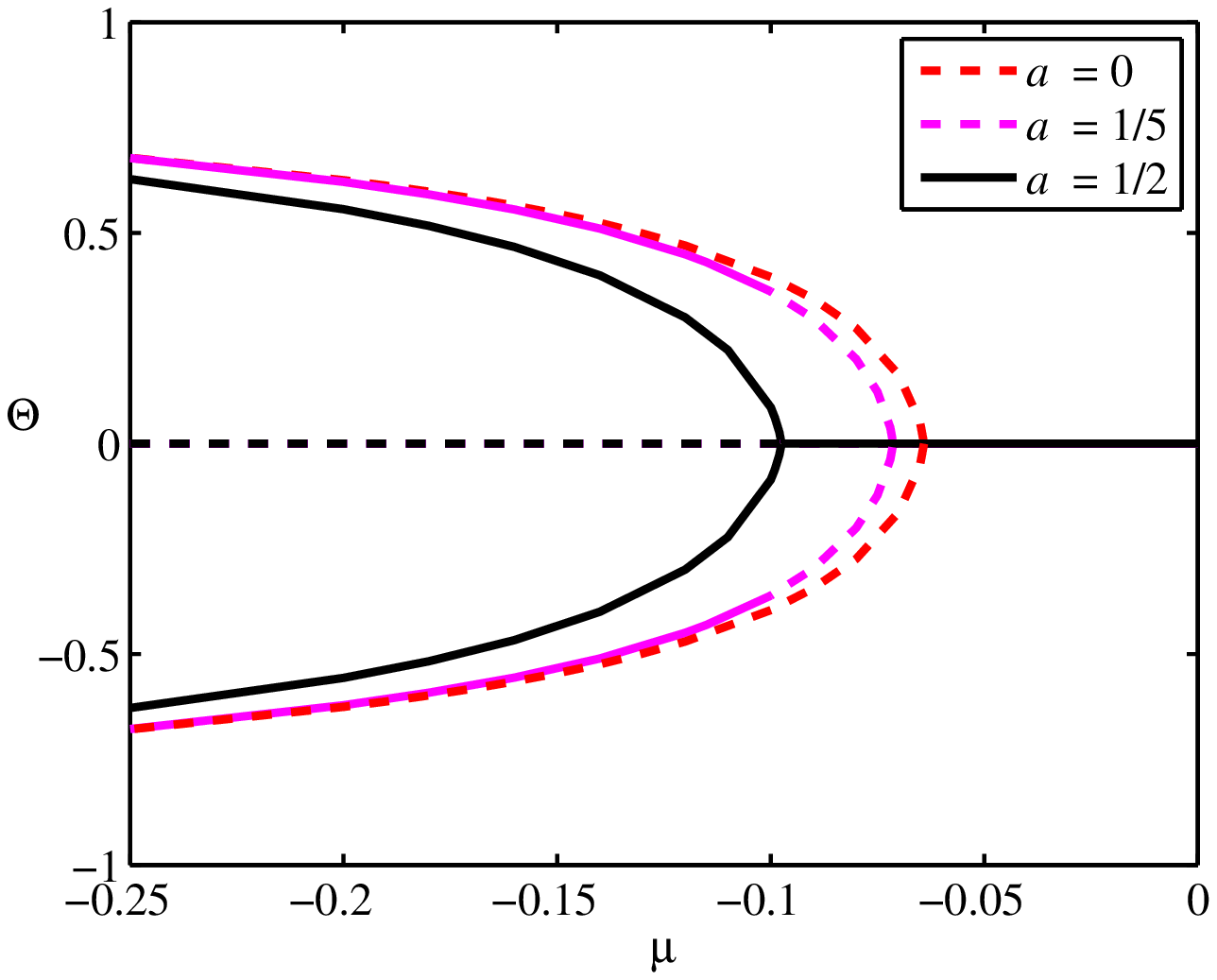}}%
\subfigure[]{\includegraphics[width=3in]{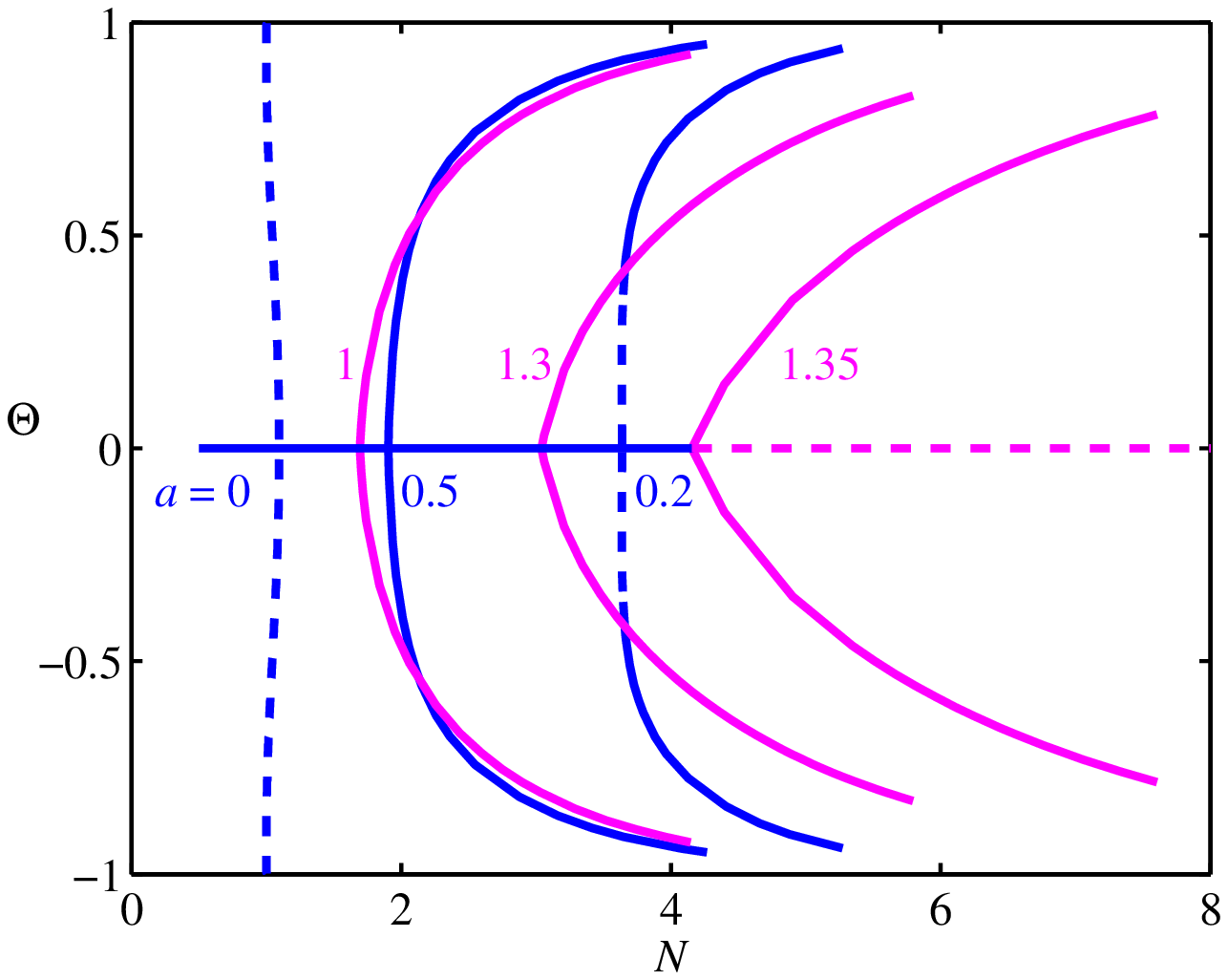}} \subfigure[]{%
\includegraphics[width=3in]{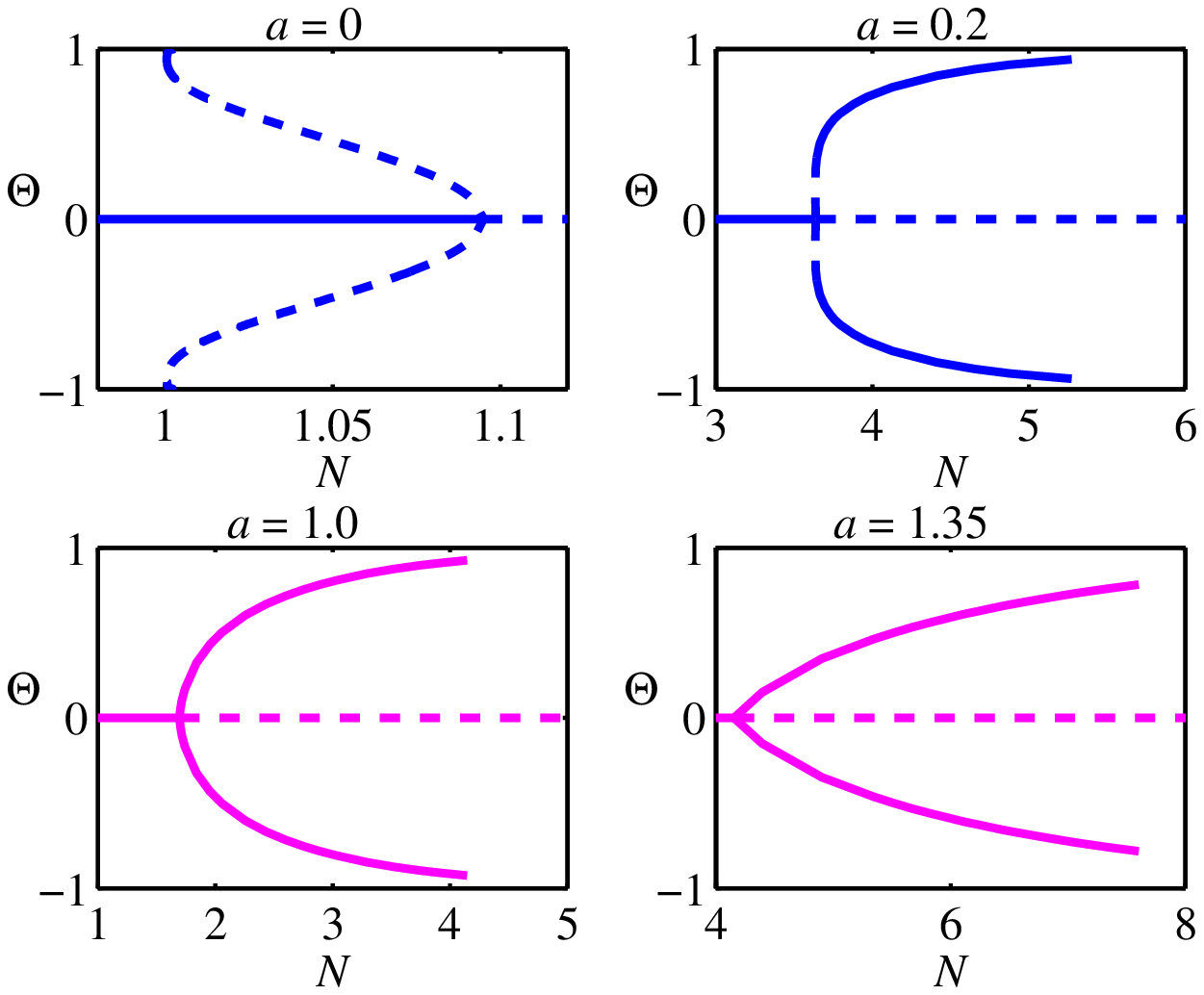}}\subfigure[]{%
\includegraphics[width=3in]{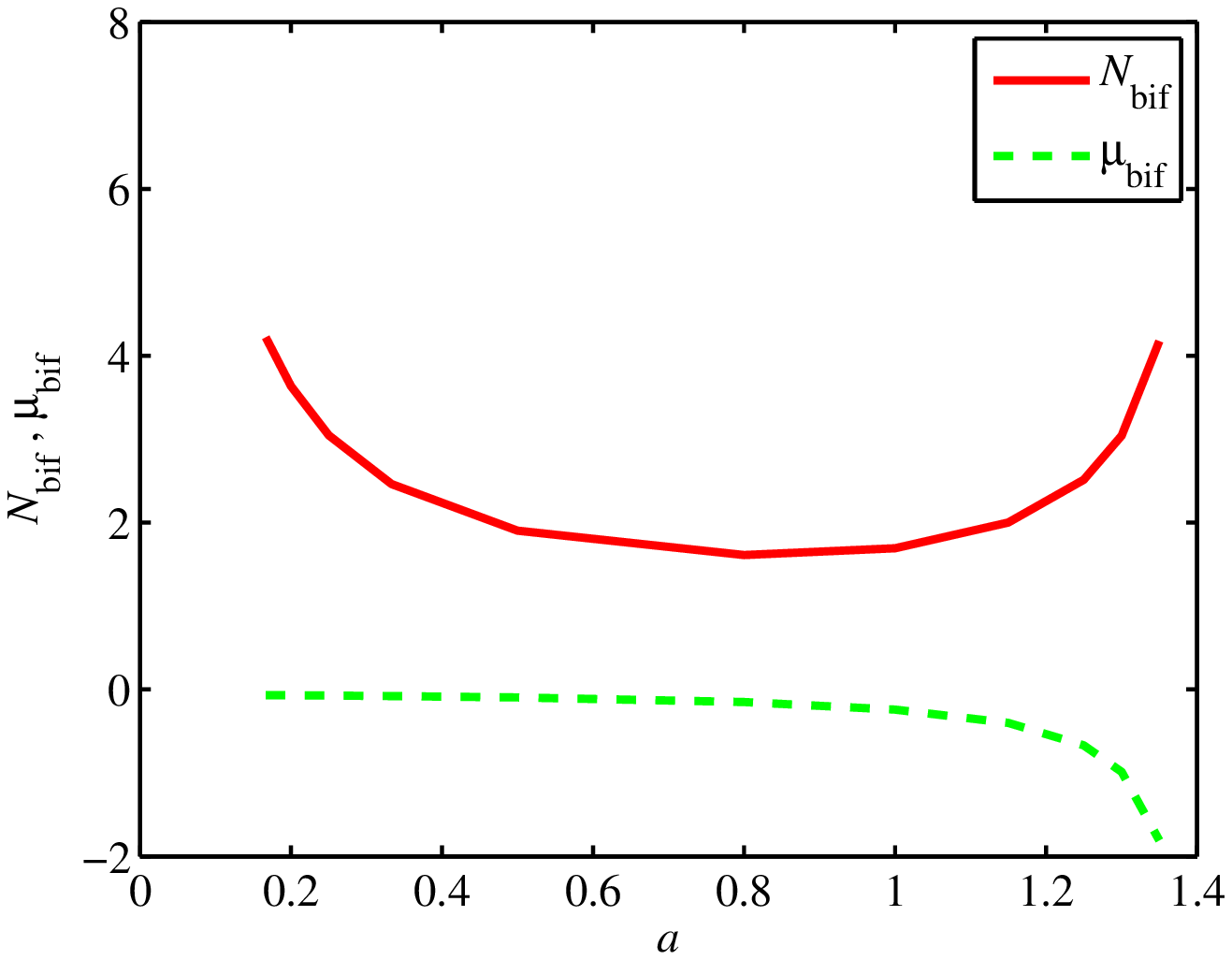}}
\caption{(Color online) Families of asymmetric states in the plane of $%
\left( \Theta ,\protect\mu \right) $ (a) and $\left( \Theta ,N\right) $ (b)
for different fixed values of nonlinear-potential parameter $a$ [recall that
asymmetry ratio $\Theta $ is defined by Eq. (\protect\ref{theta}), solution
families with $\Theta \equiv 0$ being symmetric ones; panel (b) takes into
regard the fact that symmetric states do not exist at $N<1/2$]. Plots
labeled ``$a=0$" were generated by analytical solutions (%
\protect\ref{linear}), (\protect\ref{A0B0}) and (\protect\ref{asymm}). Panel
(c) additionally displays solution branches from (b) near the bifurcation
point, and (d) shows the coordinates of the bifurcation point, $N_{\mathrm{%
bif}}$ and $\protect\mu _{\mathrm{bif}}$, as functions of $a$.
Here and in Fig. \protect\ref{Fig9} below, solid and dashed lines
depict stable and unstable solutions, respectively. Note that the
bifurcation observed in panels (b) and (c) at $a=0$ is a ``fully
backward" one: the branches of asymmetric states, which emerge at
the bifurcation point, never turn forward and, accordingly, always
remain unstable.} \label{Fig4}
\end{figure}

The analytical solution given by Eqs. (\ref{linear}), (\ref{A0B0}) and (\ref%
{asymm}) make it possible to plot the bifurcation diagrams in the planes of $%
\left( \mu ,\Theta \right) $ and $\left( N,\Theta \right) $, which are
represented by curves pertaining to $a=0$ in Figs. \ref{Fig4}(a,b,c). To
generate the diagrams, partial norms $N_{\pm }$ in expression (\ref{theta})
for the asymmetric solutions were computed numerically [analytical
expressions for them are available, but they are very messy, cf. Eqs. (\ref%
{+-}) for the symmetric and antisymmetric states].

A salient peculiarity of the SSB bifurcation for $a=0$, evident in Fig. \ref%
{Fig4}(c), is its \textit{subcritical} character, which means that
the branches of asymmetric solutions emerge at the bifurcation
point as unstable ones, and go in the \textit{backward} direction.
A subcritical bifurcation also occurs in the above-mentioned model
of the dual-core nonlinear fibers \cite{dual-core-fiber}, but in
that case the asymmetric branches quickly turn in the forward
direction, getting stabilized at the turning point. A remarkable
feature of the present model with $a=0$ is that this does not
happen, i.e., the bifurcation in this model may be called a
``fully backward" one: the branches of the asymmetric solutions
keep going backward up to the limit of $\Theta =1$, which
corresponds to the asymmetric solutions with $\mu =-$ $\infty $
(and $N=1$, as shown in the following subsection). Indeed, $\Theta
(\mu =-\infty )=1$ follows from the fact that the amplitude
appertaining to the lower sign between the radicals in Eq.
(\ref{asymm}) vanishes in the limit of $\mu \rightarrow -\infty $.

In accordance with general properties of the subcritical SSB\ bifurcation
\cite{dual-core-fiber}, the symmetric solution is expected to be stable
below the bifurcation point [at $N<N_{\mathrm{bif}}$, see Eq. (\ref{Nbif})],
and unstable above it. The asymmetric branches emerging at $N=N_{\mathrm{bif}%
}$ are unstable as long as they go backward. In the present case ($a=0$),
this means they are always unstable, as the respective branches in Figs. \ref%
{Fig4}(b,c) never turn forward. All these expectations are completely borne
out by the stability analysis performed by means of both direct simulations
and computation of stability eigenvalues, at finite but small values of $a$
(technical details of the procedure are described in the next section). In
particular, at $N>N_{\mathrm{bif}}$ the unstable symmetric state
spontaneously transforms into a strongly asymmetric breather which features
irregular oscillations, but remains robust as a whole (quite similar to an
example displayed below in Fig. \ref{NewFig3} for $a=0.7$). On the other
hand, unstable asymmetric states transform themselves into breathers which
maintain the original asymmetry of the unstable state, as shown in Fig. \ref%
{Fig3}.
\begin{figure}[h]
\centering\includegraphics[width=4in]{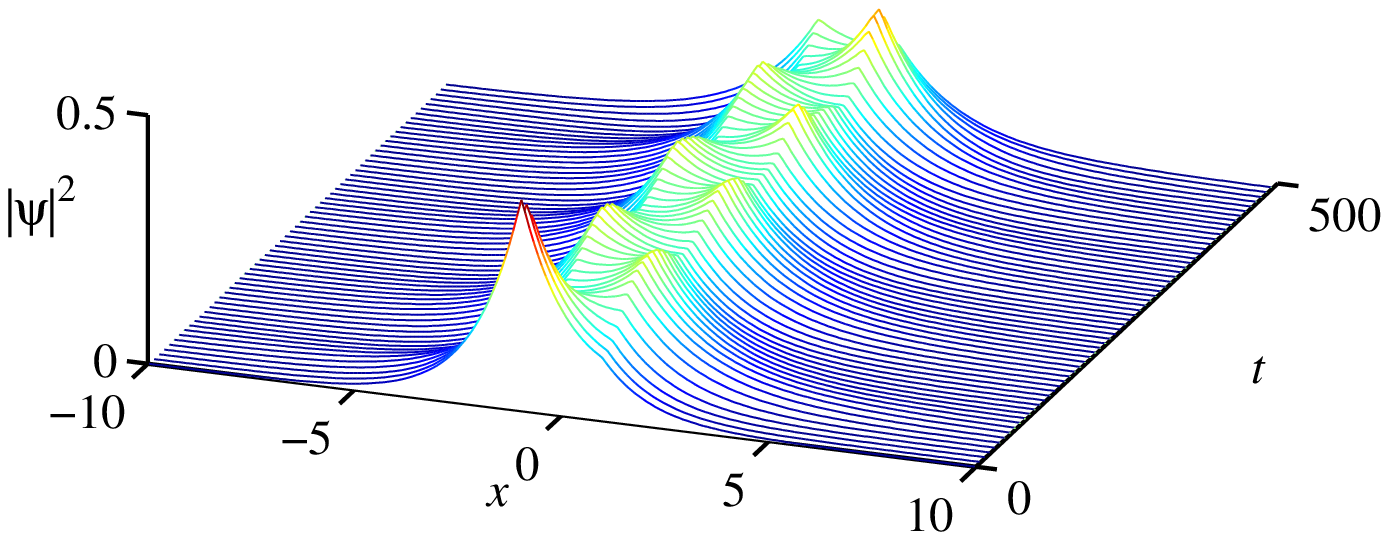}
\caption{(Color online) The evolution of an unstable asymmetric state for $%
a=1/60$ and $\protect\mu =-0.1$.}
\label{Fig3}
\end{figure}

Lastly, all antisymmetric states in the model with delta-functions are
unstable too. Their instability is similar to that shown below in Fig. \ref%
{Fig6}(b) for $a=1$, transforming them into strongly asymmetric breathers.
As shown in the next section, both asymmetric and antisymmetric states may
be stabilized in the general model, with finite $a$.

\subsection{Concluding remarks concerning the delta-function model}

The existence of the asymmetric states, i.e., the presence of the SSB effect
in the present model, can be easily explained by the consideration of the
above-mentioned limit of $\mu \rightarrow -\infty $. Indeed, in this limit,
the spatial scale of the solution, which is $\sim \left\vert \mu \right\vert
^{-1/2}$ according to Eq. (\ref{linear}), is much smaller than the
separation between the two delta-functions, $2\Lambda \equiv 2$. Therefore,
the full solution effectively splits into a superposition of those
independently supported by each delta-function in isolation. Further, it is
obvious that, for given large $|\mu |$, Eq. (\ref{phi}) with an individual
delta-function gives rise to two solutions: a trivial one, $\phi \equiv 0$,
and%
\begin{equation}
\phi _{\pm }(x)=\pm \left( 2|\mu |\right) ^{1/4}\exp \left( -\sqrt{2|\mu |}%
\left\vert \xi \right\vert \right) ,  \label{xi}
\end{equation}%
where $\xi =x+1$ or $\xi =x-1$; note that the norm of solution (\ref{xi}) is
$N=1$, for any $\mu $. The corresponding symmetric and antisymmetric states
are built, respectively, as superpositions of solutions $\phi _{+}$ (or,
equivalently, $\phi _{-}$) centered at $x=-1$ and $x=+1$, or $\phi _{-}$
centered at $x=-1$ and $\phi _{+}$ centered at $x=+1$. Asymmetric solutions
are represented, in the same limit, by a superposition of solution $\phi
_{\pm }$ centered at $x=-1$ and zero solution around $x=+1$, or vice versa.
Of course, finding the bifurcation point requires\ one to perform the
analysis of the model at finite $\mu $, as done in the analytical form above
for the case of the delta-functions, and will be done in a numerical form
below for the general case of finite $a$ in Eq. (\ref{g}).

It is relevant to compare the above exact results with those which
can be easily obtained in the linear counterpart of the model,
i.e., the one with the DWP based on the set of two
delta-functions; as mentioned above, such a linear potential may
be a plausible ingredient of a more general model, relevant to the
description of NDWP settings in optics. The
stationary version of the linear equation reduces to%
\begin{equation}
\mu \phi +(1/2)\phi ^{\prime \prime }+\epsilon \left[ \delta (x+1)+\delta
(x-1)\right] \phi =0,  \label{linear-delta}
\end{equation}%
with constant $\epsilon >0$. Symmetric solutions, which must be continuous
at $\left\vert x\right\vert =1$, are sought for as
\begin{equation}
\phi (x)=\left\{
\begin{array}{c}
\exp \left( -\sqrt{2\left\vert \mu \right\vert }\left( |x|-1\right) \right)
,~\mathrm{at}~|x|~>1, \\
\mathrm{sech}\left( \sqrt{2\left\vert \mu \right\vert }\right) \cosh \left(
\sqrt{2\left\vert \mu \right\vert }x\right) ,~\mathrm{at}~|x|~<1,%
\end{array}%
\right. ,  \label{symmetric}
\end{equation}%
cf. Eq. (\ref{linear}). The jump condition, $\Delta \left( \phi ^{\prime
}\right) |_{x=\pm 1}=-2\epsilon \phi (x=\pm 1)$, yields equation $\sqrt{%
2\left\vert \mu \right\vert }\left[ 1+\tanh \left( \sqrt{2\left\vert \mu
\right\vert }\right) \right] =2\epsilon $, which has a single solution for $%
|\mu |$ at any positive $\epsilon $, i.e., the linear model always supports
exactly one symmetric state.

Antisymmetric solutions are sought for as%
\begin{equation}
\phi (x)=\left\{
\begin{array}{c}
\mathrm{sgn}(x)\cdot \exp \left( -\sqrt{2\left\vert \mu \right\vert }\left(
|x|-1\right) \right) ,~\mathrm{at}~|x|~>1, \\
\mathrm{cosech}\left( \sqrt{2\left\vert \mu \right\vert }\right) \sinh
\left( \sqrt{2\left\vert \mu \right\vert }x\right) ,~\mathrm{at}~|x|~<1,%
\end{array}%
\right. ,  \notag
\end{equation}%
and the respective jump condition takes the form of $\sqrt{2\left\vert \mu
\right\vert }\left[ 1+\coth \left( \sqrt{2\left\vert \mu \right\vert }%
\right) \right] =2\epsilon $. The latter equation has no solutions for $%
\epsilon <1/2$, and exactly one solution for $\epsilon >1/2$. Thus,
symmetric solution (\ref{symmetric}) is the single state in the linear DWP
model at $\epsilon <1/2$, while at $\epsilon >1/2$ the linear model supports
precisely two states, symmetric and antisymmetric ones.

Lastly, the combined model, which includes both the linear potential and its
nonlinear counterpart (that may be self-attractive, as above, or\emph{\
self-repulsive} too, in this case), is also solvable in the case when these
features are based on the pair of delta-functions. The combined model is
described by the following stationary equation, cf. Eqs. (\ref{phi}) and (%
\ref{linear-delta}):%
\begin{equation}
\mu \phi +(1/2)\phi ^{\prime \prime }+\left[ \delta (x+1)+\delta (x-1)\right]
\left( \epsilon \phi +\sigma \phi ^{3}\right) =0,  \label{combi}
\end{equation}%
where $\sigma =+1$ and $-1$ corresponds to the nonlinear
attraction and repulsion, respectively. In particular, the
bifurcation occurs only on the branch of the symmetric solutions
in the case of $\sigma =+1$, and only on the antisymmetric branch
-- in the opposite case (self-repulsion). In either case, the
value of the chemical potential ($\mu <0$) at the symmetry- or
antisymmetry-breaking bifurcation is determined by the following
transcendental equation:%
\begin{equation}
\sqrt{2\left\vert \mu \right\vert }\frac{1-2\sigma e^{-2\sqrt{2\left\vert
\mu \right\vert }}}{1-e^{-4\sqrt{2\left\vert \mu \right\vert }}}=\epsilon
\label{transc}
\end{equation}%
[for $\epsilon =0$ and $\sigma =+1$, it reduces to Eq. (\ref{bif})]. With $%
\sigma =+1$, Eq. (\ref{transc}) has exactly one solution for any $\epsilon
\geq -1/4$ and no solutions for $\epsilon <-1/4$ (negative $\epsilon $
corresponds to competition between the repulsive linear potential and
attractive nonlinear pseudopotential). With $\sigma =-1$, Eq. (\ref{transc})
has exactly one solution for any $\epsilon >3/4$, and no solutions for $%
\epsilon <3/4$. Analysis of the stability of states found in the
combined model is beyond the scope of this work, and in the rest
of the paper we consider the model without the linear potential.

\section{Numerical results for the general model}

\subsection{Numerical methods}

To construct localized solutions to stationary equation (\ref{phi2}) of the
symmetric, asymmetric and antisymmetric types, the Newton iterative method
was used, starting with the following inputs:%
\begin{equation}
\left( \phi _{0}(x)\right) _{\mathrm{sym,asym}}=B~\mathrm{sech}\left(
2\left( x+1\right) \right) +A~\mathrm{sech}\left( 2\left( x-1\right) \right)
,  \notag
\end{equation}%
\begin{equation}
\left( \phi _{0}(x)\right) _{\mathrm{antisym}}=A~\mathrm{sech}\left(
x\right) \sin \left( x\right) ,  \notag
\end{equation}%
with $B=A$ for symmetric solutions. Then, as mentioned above, numerical
analysis of the stability of the stationary solutions was performed in two
different ways: first, by means of direct simulations of the evolution of
slightly perturbed solutions, and then through computation of (in)stability
eigenvalues for modes of small perturbations. In the former case, the
stability was tested by adding arbitrary perturbations to the initial state,
at the level of $\sim 1\%$ of the amplitude (in particular, care was taken
to test effects of perturbations whose symmetry is different from that of
the stationary state, such as antisymmetric perturbations added to symmetric
states, and vice versa).

For the computation of eigenvalues, perturbed solutions were looked for as
\begin{equation}
\psi (x,t)=e^{-i\mu t}\left\{ \phi (x)+\eta \left[ u\left( x\right)
e^{i\lambda t}+v^{\ast }\left( x\right) e^{i\lambda ^{\ast }t}\right]
\right\} ,  \label{pert}
\end{equation}%
where $\phi (x)$ is a stationary solution to Eq. (\ref{phi2}) with chemical
potential $\mu $, while $u$ and $v$ are components of a perturbation mode
with an infinitesimal amplitude $\eta $, pertaining to instability growth
rate $\lambda \equiv \lambda _{\mathrm{r}}+i\lambda _{\mathrm{i}}$. The
substitution of expression (\ref{pert}) into Eq. (\ref{psi2}) and
linearization lead to the eigenvalue problem based on the following
equations:

\begin{equation}
\left(
\begin{array}{cc}
-\frac{1}{2}\frac{d^{2}}{dx^{2}}-\mu +2g\left( x\right) \phi ^{2}(x) &
g\left( x\right) \phi ^{2}(x) \\
-g\left( x\right) \phi ^{2}(x) & +\frac{1}{2}\frac{d^{2}}{dx^{2}}+\mu
-2g\left( x\right) \phi ^{2}(x)%
\end{array}%
\right) \left(
\begin{array}{c}
u \\
v%
\end{array}%
\right) =\lambda \left(
\begin{array}{c}
u \\
v%
\end{array}%
\right) .  \label{eigen}
\end{equation}%
The underlying solution, $\phi (x)$, is stable if all eigenvalues associated
with it have $\lambda _{\mathrm{i}}=0$. Equations (\ref{eigen}) were solved
numerically with the help of a finite-difference method. Conclusions
concerning the stability of the patterns, drawn from direct simulations,
always complied with results produced by the computation of eigenvalue.

\subsection{Results}

The first significant change against the results reported above for the
model with the delta-functions ($a=0$), which happens with the increase of $%
a $, is quick stabilization of asymmetric states with larger values of the
norm, while ones with smaller $N$ remain unstable, originally. At $a>0.2$,
the symmetric states are stable for all values of $N$ at which they exist.
Another notable feature of the bifurcation diagrams at finite $a$,
demonstrated by Figs. \ref{Fig4}(b) and \ref{Fig4}(d), is that the norm at
which the SSB bifurcation takes place, $N_{\mathrm{bif}}$, first decreases
with the growth of $a$ from small values up to $a\approx 0.8$, and then
increases with the further growth of $a$.

\begin{figure}[h]
\centering\subfigure[]{\includegraphics[width=3in]{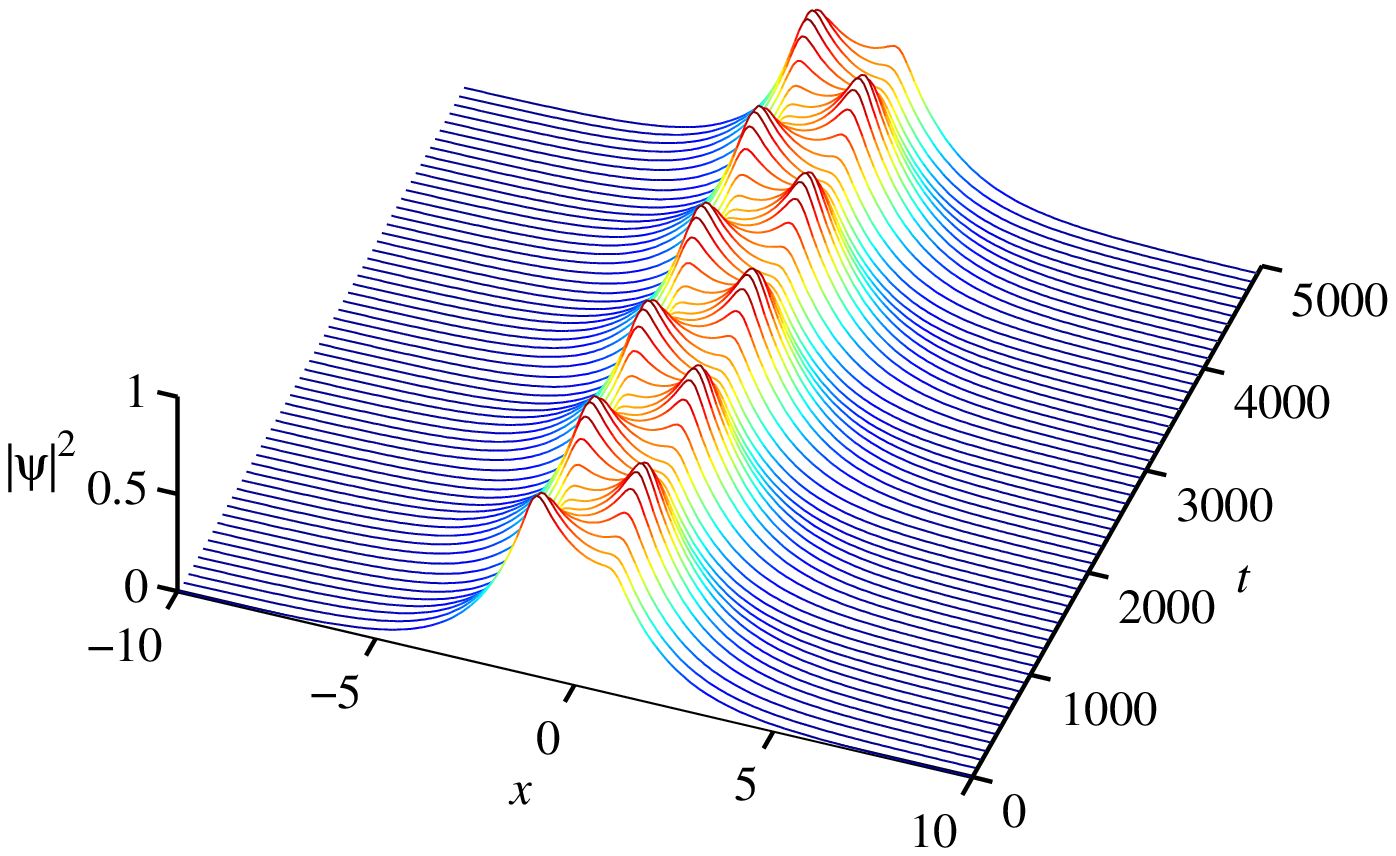}}%
\subfigure[]{\includegraphics[width=3in]{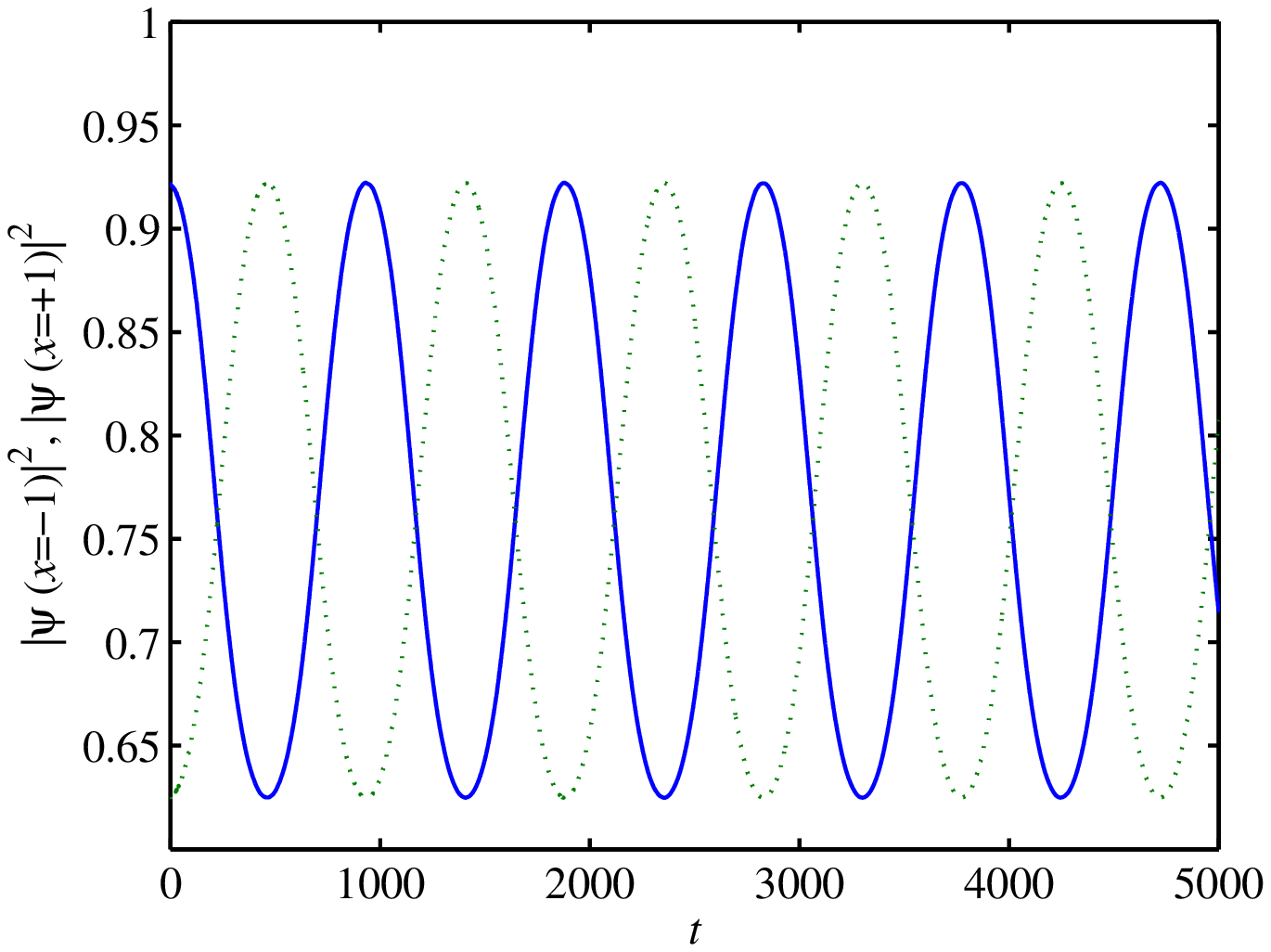}}
\caption{(Color online) (a) The evolution of a weakly unstable asymmetric
state with $a=0.2$ for $\protect\mu =-0.075$ (very close to the bifurcation
point and stabilization threshold). (b) For the same case, solid and dotted
curves show densities $\left\vert \protect\psi \left( x=-1\right)
\right\vert ^{2}$ and $\left\vert \protect\psi \left( x=+1\right)
\right\vert ^{2}$ as functions of time.}
\label{Fig5}
\end{figure}

Close to the their stabilization threshold (in particular, at $a=0.2$),
asymmetric states with a smaller norm, which are still unstable, demonstrate
a scenario of the instability development different from what was shown by
their counterpart in Fig. \ref{Fig3} in the case of very small $a$. Namely,
slow regular oscillations, observed in Fig. \ref{Fig5} in this case, imply a
\textit{dynamical re-symmetrization} of the unstable asymmetric state.
Indeed, densities $\left\vert \psi \left( x\right) \right\vert ^{2}$, taken
at points $x=\pm 1$, perform identical periodic oscillations, with a phase
shift of $\pi $ between them, as shown in \ref{Fig5}(b).

The stabilization of the asymmetric states at small finite values of $a$ is
explained by the change in the character of the SSB bifurcation: at $a\neq 0$%
, there appear turning points on branches of asymmetric solutions in the
bifurcation diagram, cf. Fig. \ref{Fig4}(c). Past the turning point, the
branch goes forward as a \emph{stable} one. In fact, Fig. \ref{Fig4}(c)
demonstrates a quick transformation, with the increase of $a$, of the
subcritical bifurcation into a supercritical one. When the bifurcation is
supercritical, branches of the asymmetric solutions emerge as stable ones at
the bifurcation point, and immediately go forward.

We do not display the quick transition from the sub- to supercritical
bifurcation in full detail, as it actually happens at very small $a$, in the
range of $a\lesssim 0.1$. The physical estimates given in Section II suggest
that so small values of the scaled width of the nonlinear-potential wells
correspond to physical widths $\lesssim 1$ $\mu $m. It seems doubtful that
the Feshbach-resonance technique would allow one to create a strong local
inhomogeneity of the scattering length on such a small scale (nevertheless,
the exact analytical solutions obtained for $a=0$, which provide clear clues
for the understanding of the general model, are definitely relevant). An
additional problem impeding the full analysis of the case of very small $a$
is that, in this case, the accumulation of systematic numerical results
requires very heavy simulations, as the stepsize of the spatial grid must be
made much smaller than $a$.

\begin{figure}[h]
\centering\includegraphics[width=4in]{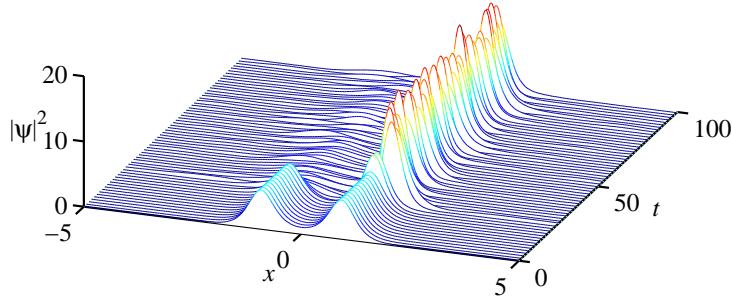}
\caption{The evolution of an unstable symmetric state, at $a=0.7$, $\protect%
\mu =-2.677$ and $N=10$.}
\label{NewFig3}
\end{figure}

Above the bifurcation point, symmetric states found at finite $a$
demonstrate the familiar SSB instability, spontaneously transforming
themselves into slightly nonstationary robust modes (breathers), quite close
in their shape to respective stable asymmetric solitons. A typical example
of this transformation in displayed in Fig. \ref{NewFig3}.

\begin{figure}[h]
\centering\subfigure[]{\includegraphics[width=3in]{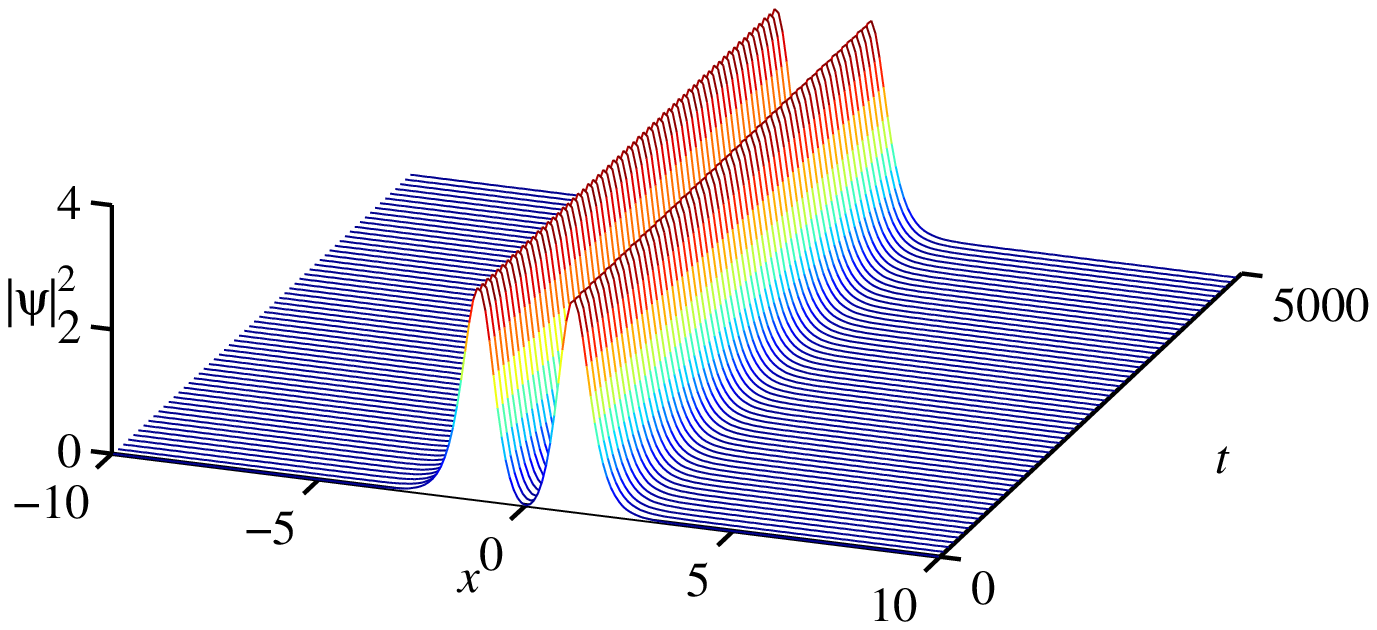}}%
\subfigure[]{\includegraphics[width=3in]{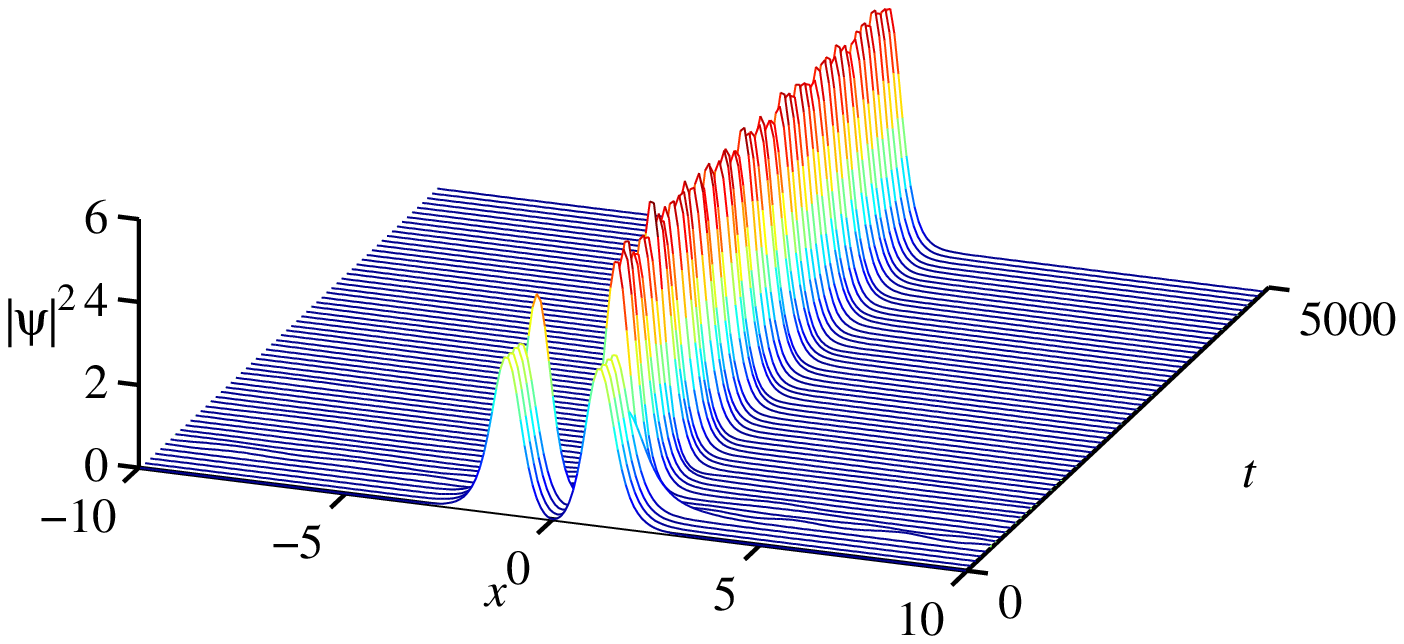}}
\caption{(Color online) (a) An example of a stable antisymmetric state with $%
a=1$, for $\protect\mu =-1.3$ and $N=6.7137$. (b) The evolution of an
unstable antisymmetric state, also with $a=1$, but for $\protect\mu =-1.5$
and $N=7.1273$.}
\label{Fig6}
\end{figure}

\begin{figure}[h]
\centering\subfigure[]{\includegraphics[width=2.35in]{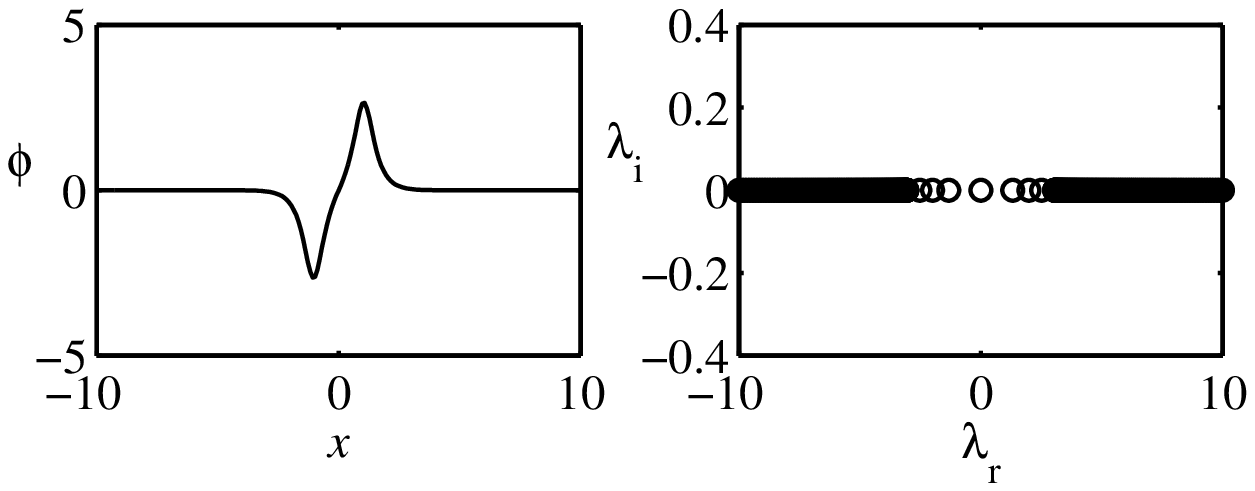}}%
\subfigure[]{\includegraphics[width=2.35in]{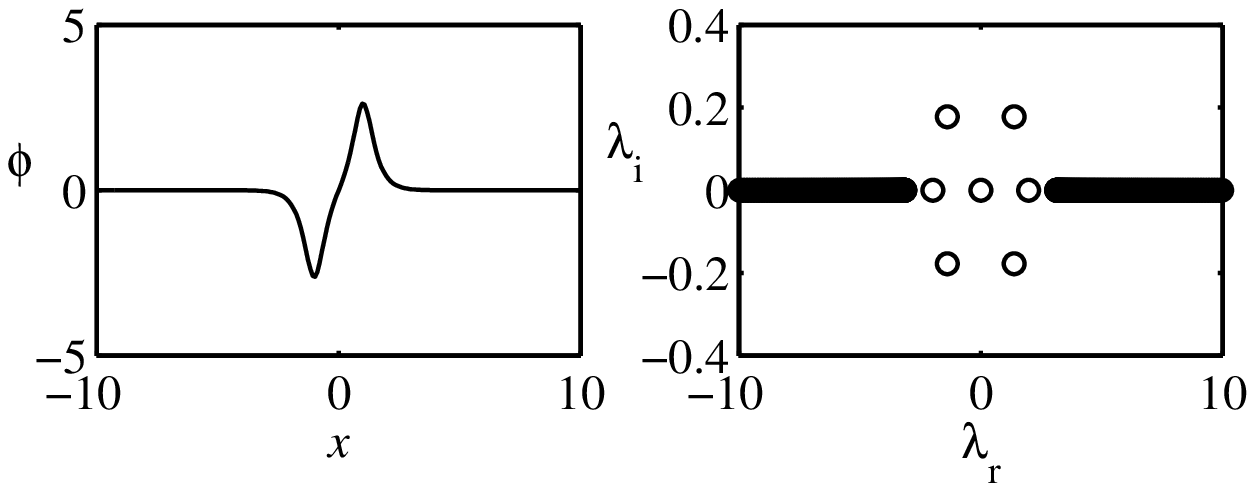}}%
\subfigure[]{\includegraphics[width=2.35in]{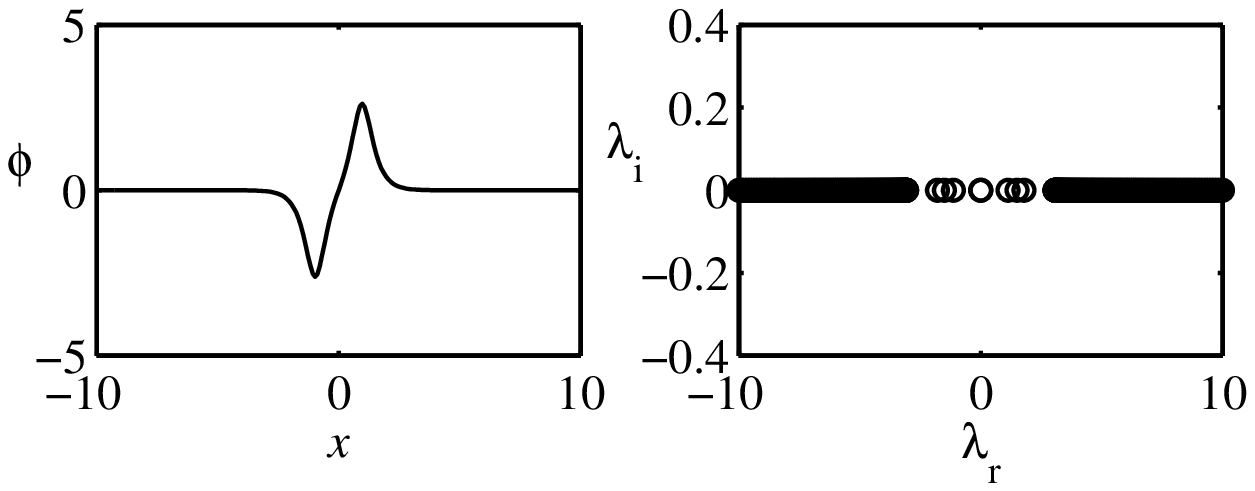}}
\caption{Examples of stable (a,c) and unstable (b) antisymmetric states,
found with a fixed norm, $N=10$. In each subplot, the left and right panels
show, severally, profiles of the stationary states and spectral planes of
the (in)stability eigenvalues. Parameters are (a) $a=0.80$ and $\protect\mu %
=-3.05$; (b) $a=1.11$ and $\protect\mu =-3.14$; (c) $a=1.30$ and $\protect%
\mu =-3.07$.}
\label{NewFig2}
\end{figure}

As concerns antisymmetric solutions, both stable and unstable ones have been
found at finite $a$, as illustrated by Figs. \ref{Fig6} and \ref{NewFig2}.
Panel (b) of the former figure shows that the density profile of unstable
antisymmetric states evolves from the double-peak pattern into an asymmetric
single-peak one, which features persistent intrinsic oscillations. This
outcome of the instability development complies with the fact that the
instability of the antisymmetric states is oscillatory, being accounted for
by a quartet of eigenvalues, as seen in Fig. \ref{NewFig2}(b). In other
words, the transition from stable to unstable antisymmetric states may be
considered as the \textit{Hamiltonian Hopf bifurcation} \cite{HH}.

\begin{figure}[h]
\centering\includegraphics[width=3.5in]{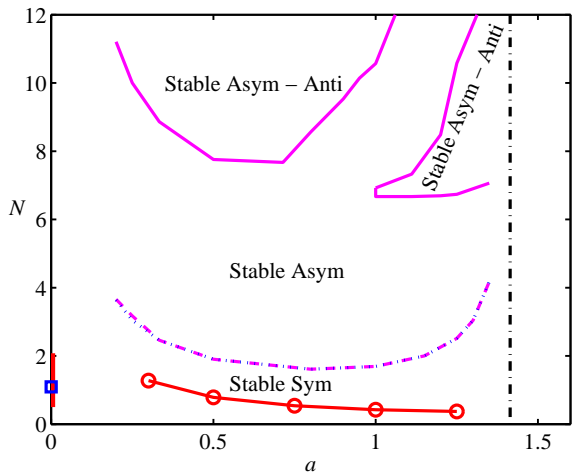}
\caption{(Color online) Stability and existence borders for
symmetric, asymmetric and antisymmetric states in the plane of
$\left( a,N\right) $. The vertical dashed-dotted line corresponds
to $a=\protect\sqrt{2}$, beyond which the double-well structure in
Eq. (\protect\ref{g}) turns into a single-well one. The lower
dashed-dotted line depicts the symmetry-breaking bifurcation, with
stable asymmetric solutions existing above it. The chain of
circles in the bottom of the parameter plane designates the
minimum norm necessary for the existence of symmetric states,
which are stable between the existence border and the bifurcation
line. Label ``Stable Asym - Anti" marks regions of the
bistability, where both asymmetric and antisymmetric states are
stable.} \label{Fig8}
\end{figure}

Figure \ref{Fig8} displays a combined diagram in the plane of the norm of
the solution and width of the nonlinear potential wells, $N$ and $a$, which
summarizes the existence and stability results for the states of all the
three types -- symmetric, asymmetric and antisymmetric ones. The
dashed-dotted line in the figure designates the symmetry-breaking
bifurcation. Solely symmetric states exist below this line (they are stable
in that region), and stable asymmetric states exist above the line, where
the symmetric ones are unstable. Solid curves in Fig. \ref{Fig8} depict
stability borders of antisymmetric solutions.

For the reasons explained above, the region of very small values
of $a$, where the ``quick" stabilization of asymmetric states
takes place, is not included. However, the region of the existence
of the analytical symmetric and antisymmetric solutions in the
model with delta-functions ($a=0$), and the respective bifurcation
point, as given by Eq. (\ref{Nbif}), are shown by the bold
vertical segment and square-marked dot on the axis of $a=0$
(recall that the exact asymmetric solutions are unstable above the
bifurcation point in the model with $a=0$).

Because the modulation profile (\ref{g}) does not feature the double-well
structure for $a\geq \sqrt{2}$ (see Fig. \ref{Fig1}), the SSB\ bifurcation
tends to disappear as $a$ approaches $\sqrt{2}$. Actual results are included
in Fig. \ref{Fig8} for $a\leq 1.35,$ as the convergence of the numerical
scheme becomes poor for values of $a$ still closer to $\sqrt{2}$.

The chain of circles in Figure \ref{Fig8} designates the threshold (minimum
norm), $N_{\min }$, necessary for the existence of symmetric states in the
model. As mentioned above, in the case of $a=0$ the exact threshold is $%
N_{\min }(a=0)=1/2$, and it is observed in Fig. \ref{Fig8} that the
threshold remains in the ballpark of this value at finite $a$, which can be
easily explained. Indeed, the minimum of $N$ is attained at $\mu \rightarrow
-0$, in which limit the spatial scale of the wave function, $\sim 1/\sqrt{%
2|\mu |}$, is much larger than the size of the NDWP structure, $2\Lambda
\equiv 2$. Thus, from the viewpoint of this weakly localized wave function,
modulation pattern (\ref{g}) looks like $2\delta (x)$. In the corresponding
approximation, the wave function takes the form of expression (\ref{xi})
divided by $\sqrt{2}$, and the respective norm is, indeed, $1/2$.

\begin{figure}[h]
\centering\includegraphics[width=3.5in]{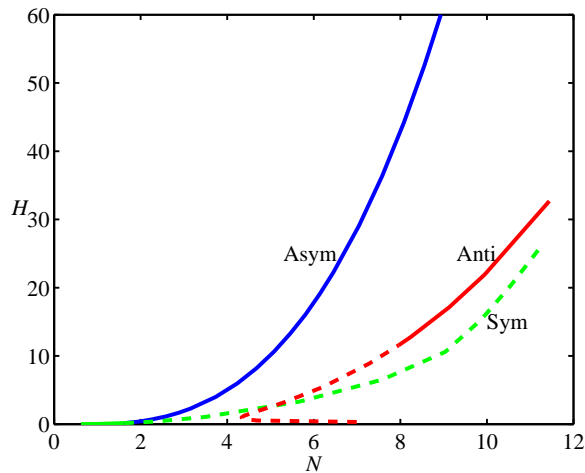}
\caption{(color online) The Hamiltonian versus the norm for solutions of
different types, with fixed $a=0.75$.}
\label{Fig9}
\end{figure}

For $N<N_{\min }$, the condensate confined to the trap of large length $L$
(i.e., in the thermodynamic limit) will tend to form a quasi-uniform nearly
linear state, with $\phi (x)=\sqrt{N/L}$. As follows from Eqs. (\ref{N}) and
(\ref{g}), the energy of the small-amplitude uniform state is%
\begin{equation}
H_{0}\approx -N^{2}/L^{4}.  \label{GS}
\end{equation}%
In fact, this state realizes a minimum of the energy (cf. Fig. \ref{Fig9}),
i.e., the system's ground state. Nevertheless, a well-known fact is that
dynamically stable localized states \emph{different} from the ground state,
such as the above-mentioned gap solitons in the repulsive condensate \cite%
{Heidelberg}, or their broader counterparts, in the form of the so-called
gap-waves \cite{gap-wave}, can be created in the experiment.

Another notable feature observed in Fig. \ref{Fig8} is the \textit{%
bistability}, i.e., the coexistence of stable asymmetric and antisymmetric
states above the stability border of the latter state. In fact, the
bistability always takes place when antisymmetric states are stable. It is
interesting too that the stability area for the antisymmetric states
consists of two separate regions. Finally, it is relevant to mention that,
as well as in the analytically solvable model with the delta-functions ($%
a\rightarrow 0$), the antisymmetric states never undergo a bifurcation at
finite $a$.

We stress that the stability borders displayed in Fig. \ref{Fig8} were
identified by means of direct simulations and the computation of stability
eigenvalues, both methods yielding identical results. In particular, the
sets of eigenvalues displayed in Fig. \ref{NewFig2} clearly confirm the
presence of two disjoint stability areas for antisymmetric states.

In the case of the bistability involving the asymmetric and antisymmetric
states, it is interesting to compare their energies (values of the
Hamiltonian). To this end, Fig. \ref{Fig9} displays a typical example of the
dependence of $H$ on norm $N$. The situation observed in this figure is also
true in the general case: stable antisymmetric states realize smaller values
of $H$ than the asymmetric counterparts coexisting with them. However, as
argued above, dynamically stable states can be created in the experiment
even if their energy is higher than in some competing states. In particular,
Fig. \ref{NewFig3} demonstrates that an unstable symmetric state definitely
self-traps into an asymmetric robust breather (which is close to a stable
stationary solution), despite the fact that a stable antisymmetric state
exists at the same values of $N=10$ and $a=0.7$, as seen from Fig. \ref{Fig8}%
.

Note that all curves in Fig. \ref{Fig9} start from finite threshold values
of $N$ corresponding, as said above, to the minimum norm ($N_{\min }$)
necessary for the existence of the respective states. In particular, for the
branch of symmetric solutions, $N_{\min }$ is close to $1/2$, as argued
above (cf. the existence border in the bottom of Fig. \ref{Fig8}), while the
asymmetric branch originates at the bifurcation point (in agreement with the
location of the respective dashed-dotted line in Fig. \ref{Fig8}), at which
the symmetric solution loses its stability. The branch of antisymmetric
solutions features a fold in Fig. \ref{Fig9} (in the region where these
solutions are unstable), which is similar to the the above-mentioned fact
that dependence $N(\mu )$ in Eq. (\ref{+-}) for the unstable exact
antisymmetric states has a minimum, $\left( N_{\mathrm{antisym}}\right)
_{\min }\approx 1.84$ at $\mu \approx -0.58$. If replotted in terms of $H$
and $N$, Eq. (\ref{+-}) features a similar fold, at $N=\left( N_{\mathrm{%
antisym}}\right) _{\min }$.

\section{Conclusion}

In this work, we have introduced a model of the nonlinear double-well
potential (NDWP), alias a double-well pseudopotential, which can be created
in BEC, by means of the spatially inhomogeneous Feshbach resonance, and also
in nonlinear optics. The model provides for a previoulsy unexplored setting
in which effects of the spontaneous symmetry breaking (SSB) can be studied.

In the limit when each potential well is induced by the
delta-function, full analytical solutions were obtained for
symmetric, antisymmetric and asymmetric states. The symmetric
states are stable in that case up to the symmetry-breaking
bifurcation point, but beyond the bifurcation both symmetric and
emergent asymmetric states are unstable. In particular, the
asymmetric configurations transform themselves into breathers. The
instability of all the stationary asymmetric states in the model
with the delta-functions is explained by fact that the respective
SSB bifurcation is of a ``fully backward" type, with branches of
the asymmetric solutions never turning forward. All antisymmetric
states are unstable too, in this limit form of the model.

The increase of the width of the potential wells readily stabilizes the
asymmetric states, which concurs with the change of the character of the SSB
bifurcation from sub- to supercritical. Close to the stabilization border,
unstable asymmetric states develop slow intrinsic oscillations, featuring
effective dynamical re-symmetrization. Antisymmetric states may also be
stable in the NDWP structure with a finite width of the wells, which implies
the bistability between asymmetric and antisymmetric states. The symmetric
states exist above a finite threshold, in terms of the norm (number of atoms
in the condensate), and they develop the usual SSB instability above the
bifurcation point. A simple explanation to the existence threshold was
given, and an integrated diagram for the existence and stability of the
trapped states of all the three types has been produced.

The analysis presented in this work suggests new experiments in the
matter-wave and nonlinear-optical settings. The analysis can also be
developed in other directions. In particular, it may be interesting to study
a two-dimensional nonlinear-DWP configuration. In the 2D space, a triangular
configuration with three nonlinear (pseudo-)potential wells may be
considered too.

The work of T.M. is supported, in a part, by a postdoctoral fellowship from
the Pikovsky-Valazzi Foundation, by the Israel Science Foundation through
the Center-of-Excellence grant No. 8006/03, and by the Thailand Research
Fund under grant No. MRG5080171.

\end{document}